\documentstyle[psbox]{WORKSHOP}
\begin{document}

\title{
Testing the No-Hair Theorem with Observations of \\Astrophysical Black Holes
in the Electromagnetic Spectrum}

\author{
Dimitrios Psaltis$^{1,2,3}$ and Tim Johannsen$^{2}$
\\[12pt]  
%
$^1$ Astronomy Department, University of Arizona, 933 N.\ Cherry
Ave, Tucson, AZ 85721, USA\\
$^2$ Physics Department, University of Arizona, 1118 E.\ 4th St., 
Tucson, AZ 85721, USA\\
$^3$ Institute for Theory and Computation, Harvard-Smithsonian Center
for Astrophysics,\\ 60 Garden St., Cambridge, MA 02138, USA\\
%
{\it E-mail(DP): dpsaltis@email.arizona.edu, (TJ): timj@physics.arizona.edu}
}

\abst{The Kerr spacetime of spinning black holes is one of the most
  intriguing predictions of Einstein's theory of general
  relativity. The special role this spacetime plays in the theory of
  gravity is encapsulated in the no-hair theorem, which states that
  the Kerr metric is the only realistic black-hole solution to the
  vacuum field equations. Recent and anticipated advances in the
  observations of black holes throughout the electromagnetic spectrum
  have secured our understanding of their basic properties while
  opening up new opportunities for devising tests of the Kerr
  metric. In this paper, we argue that imaging and spectroscopic
  observations of accreting black-holes with current and future
  instruments can lead to the first direct test of the no-hair
  theorem. }

\kword{relativity --- black hole physics}

\maketitle
\thispagestyle{empty}

\section{Introduction}

According to the general relativistic no-hair theorem, the Kerr metric
is the only axisymmetric, vacuum solution to the Einstein field
equations that possesses a horizon and no time-like loops (see Heusler
1996 and references therein).  Together with the cosmic censorship
conjecture, which states that a naked singularity cannot be formed by
an astrophysical process (see, however, Shapiro et al.\ 1995), this
theorem naturally leads to the expectation that all astrophysical
objects that have been identified as black-hole candidates are indeed
described by the Kerr metric.

The Kerr nature of astrophysical black holes is a prediction that
needs to be tested observationally. It is still mathematically
possible within general relativity that these astrophysical objects
are described by a metric with a naked singularity and no horizon,
such as the Manko \& Novikov (1992) metric, violating the cosmic
censorship hypothesis. Alternatively, the massive objects we have
identified as black-hole candidates may be ultra massive ``stars''
(such as boson stars, Q stars, gravastars, etc.) supported by fields
and matter in conditions that we have not encountered in terrestrial
experiments (see the discussion in Psaltis 2006; Barcel\'o et al.\ 2008;
Narayan \& McClintock 2008).  Finally, the theory of general
relativity itself may break down in the strong-field regime, with the
more complete theory leading to a black hole solution that is not
described by the Kerr metric (e.g., Yunes \& Pretorious 2009; also
Psaltis et al.\ 2008 and Barausse \& Sotiriou 2008). Testing the
no-hair theorem with astrophysical black holes offers the unique
opportunity for both rejecting alternative interpretations of their
nature and for verifying general relativity in the strong-field regime.

The substantial improvement in observational techniques of the last
decade has led to the identification of at least four observables from
accreting black holes that depend on the spacetimes very close to
their event horizons and allow, in principle, for a test of the
no-hair theorem (see also Psaltis 2008): {\em (i)\/}~the
high-resolution images of the inner accretion flows (Doeleman et
al.\ 2008; Broderick et al.\ 2009), {\em (ii)\/}~ the relativistically
broadened iron lines in their X-ray spectra (e.g., Reynolds \& Nowak
2003; Fabian 2007; Nandra et al.\ 2007; Miller 2007), {\em
  (iii)\/}~the maxima of the thermal spectra from their accretion
disks (e.g., Shafee et al. 2006, Narayan et al.\ 2007), and {\em
  (iv)\/}~the quasi-periodic oscillations in their X-ray lightcurves
(Psaltis 2004; Remillard \& McClintock 2006).

On the theoretical front, there have also been significant recent
advances in the development of frameworks with which observations of
black holes can be used to test quantitatively the Kerr metric and
search for violations of the no-hair theorem. These involve, for
example, the expansion of the black-hole spacetime into multipoles
with coefficients that can be measured observationally (Ryan 1995), or
parametric deviations of the Schwarzschild (Collins \& Hughes 2004)
and of the Kerr spacetimes (Glampedakis \& Babak 2006; Vigeland \&
Hughes 2009) for black holes with zero or finite spins,
respectively. Although these studies focused on the emission of
gravitational waves from inspirals of compact objects onto
supermassive black holes, the basic methods they advocated can be
extended to analyze and understand observations of black holes in the
electromagnetic spectrum.

In this paper, we describe a parametric framework with which tests of
the no-hair theorem can be formulated and performed with imaging and
spectroscopic observations. We then explore the observable
implications of a violation of the no-hair theorem and discuss the
strategies with which a test of the theorem can be performed in the
near future.

\section{Parametrizing Violations of the No-Hair Theorem}

The Kerr metric is uniquely determined by only two parameters: the
mass and the spin of the black hole (we do not consider here the
unlikely possibility that an astrophysical black hole will have a net
charge). This allows us to define a formal test of the no-hair theorem,
based on the work of Ryan (1995), in the following way (see also
Collins \& Hughes 2004; Glampedakis \& Babak 2006; Gair et al.\ 2008;
Vigeland \& Hughes 2009).

We can, in principle, expand the exterior metric of any compact object
in multipoles (Geroch 1970; Hansen 1974) and use observations to
measure the coefficients of the expansion. Because of the no-hair
theorem, only two of the multipole coefficients for the spacetime of a
black hole are independent. The coefficient of the monopole is the
mass $M$ of the black hole and of the dipole is its spin $a$. All
higher-order coefficients will depend on the first two, in the
particular way dictated by the Kerr metric. Testing the no-hair
theorem requires measuring at least the coefficient of the quadrupole
$q$ and verifying whether it satisfies the Kerr relation $q=-a^2$.

Four different approaches have been explored so far for introducing
additional non-Kerr hair to the spacetimes of compact objects. Ryan
(1995) studied a general expansion of stationary, axisymmetric, and
asymptotically flat spacetimes in Geroch-Hansen multipoles. Collins \&
Hughes (2004) as well as Vigeland \& Hughes (2009) added Weyl-sector
bumps to the Schwarzschild and Kerr spacetimes. Glampedakis \& Babak
(2006) used the Hartle-Thorne metric, which is valued for slowly
spinning compact objects in general relativity, and allowed its
quadrupole moment to attain non-Kerr values. Finally, Gair et
al.\ (2008) considered a coupled set of multipole moments in the Manko
\& Novikov (1992) spacetime that depends on three parameters.

All the above approaches were developed originally in order to test
general relativity with future observations of the gravitational waves
generated during inspirals into supermassive black holes (see Hughes
2006). The calculation of the waveforms of the gravitational waves
themselves requires the solution of the time-dependent Einstein field
equations on the parametric post-Kerr background. As a result, the
validity of general relativity is assumed implicitly in the
computation of these waveforms. This is not the case, however, when
predicting observables in the electromagnetic spectrum, which can be
calculated by requiring only the validity of the equivalence principle.

As a first approach to testing the no-hair theorem, we use the
parametric post-Kerr spacetime obtained by Glampedakis \& Babak
(2006).  This approach uses a single parameter associated to the
quadrupole moment of the spacetime to quantify potential deviations
from the Kerr metric, making it the simplest and most concise possible
avenue for testing the no-hair theorem. Moreover, the complete metric
of Glampedakis \& Babak (2006) remains a valid solution to the vacuum
Einstein field equations, allowing us to perform a self-consistent
test of the theorem and of the black-hole identification of the
compact object, within general relativity. The key drawback of this
metric is that it cannot be used in describing the exterior spacetimes
of rapidly spinning black holes. It will, of course, be optimal to
perform the tests of the no-hair theorem described below with all
four of the above formalisms in order to explore the robustness of
the results.

\begin{figure*}[t]
\centering
\psbox[xsize=8.0cm]
{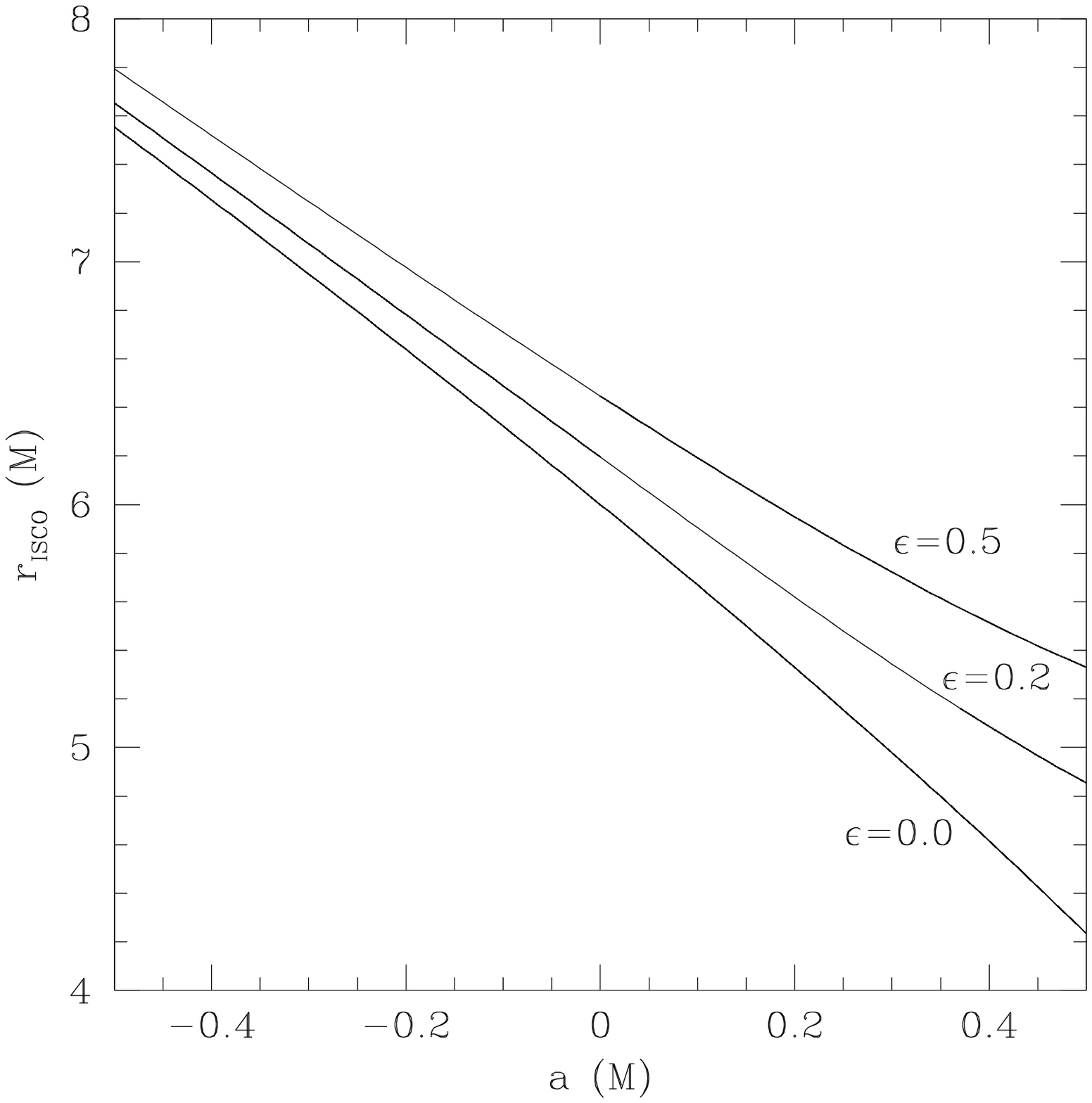}
\psbox[xsize=8cm]
{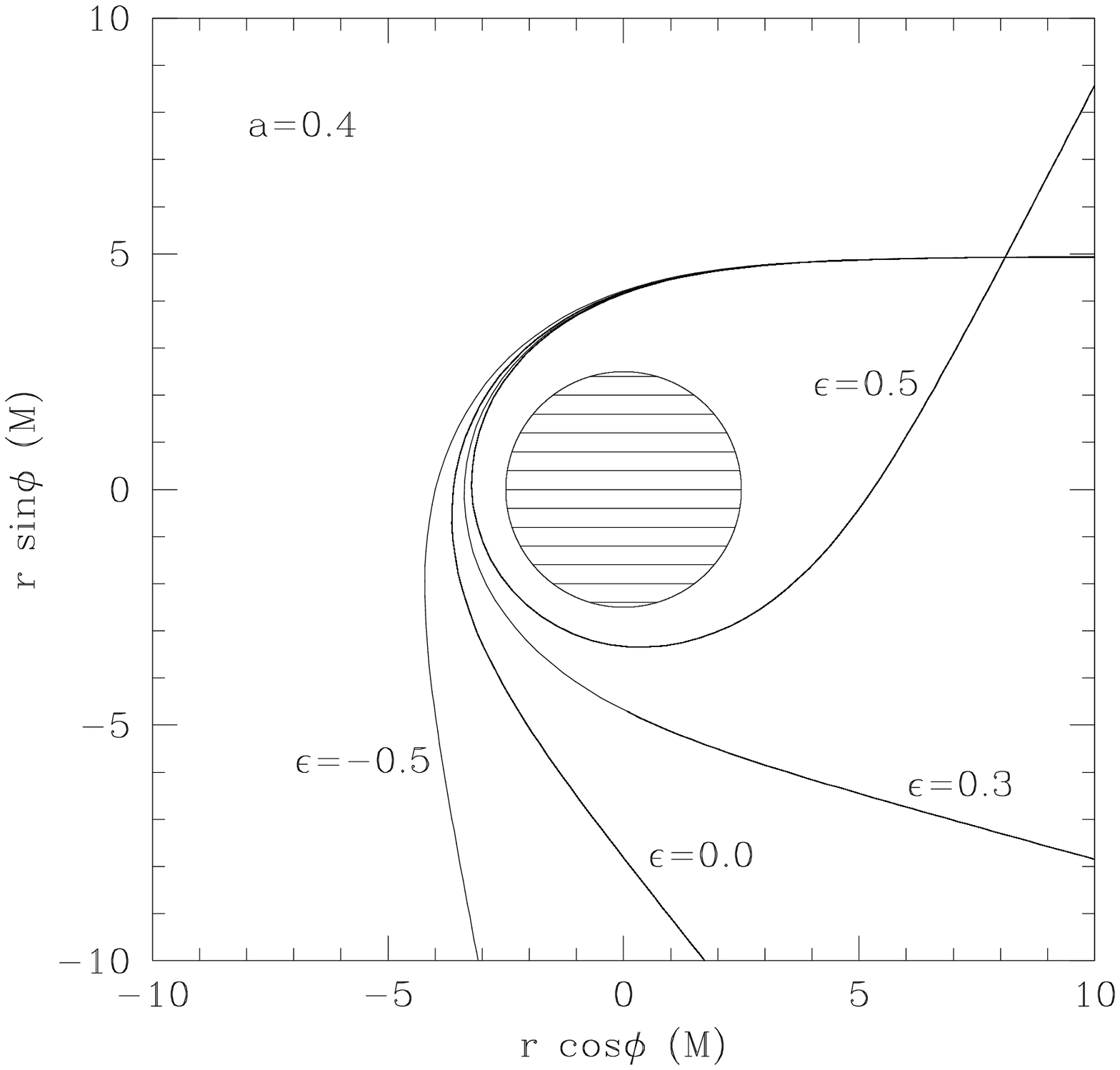}
\caption{{\em (Left)\/} The dependence of the location of the
  innermost stable circular orbit around a black hole on the
  black-hole spin $a$ and on the parameter $\epsilon$ that measures
  the degree of violation of the no-hair theorem. {\em (Right)} The
  trajectories of photons on the equatorial plane in the vicinity of
  the black-hole horizon for spacetimes with different values of the
  parameter $\epsilon$ (Johannsen \& Psaltis 2009).}
\label{fig:quadrupole}
\end{figure*}

Following Glampedakis \& Babak (2006), we start with the most
general axisymmetric spacetime of a slowly spinning compact object in
general relativity, allowing for its quadrupole moment $q$ to take
arbitrary values. We express the coefficient of the quadrupole
multipole as
\begin{equation}
q=-(a^2+\epsilon)
\end{equation}
with the parameter $\epsilon$ measuring the degree of violation of the
no-hair theorem and all the multipole coefficients expressed in
geometric units. We then study the trajectories of photons and
particles in this spacetime and identify the implications for various
observables of the presence of a non-Kerr quadrupole in the spacetime
of a black hole.

\section{The observational appearance of black holes that violate
the no-hair theorem}

We explored in detail the Glampedakis \& Babak (2006) metric in
Johannsen \& Psaltis (2009), addressing its potential for testing the
no-hair theorem with observations in the electromagnetic spectrum. We
identified a number of properties of the spacetimes that are
significantly affected by the presence of a non-Kerr quadrupole, with
observable consequences.

\noindent {\em (i) The location of the innermost stable circular orbit
  (ISCO)\/}. This is shown in Figure~\ref{fig:quadrupole}, as a
function of the spin of the black hole, for different values of the
parameter $\epsilon$.  The location of the maximum emission from the
accretion flow around a black hole is expected to be very close to
that of the ISCO (see, e.g., Krolik \& Hawley 2002 and references
therein). As a result, the maximum temperatures of geometrically thin
accretion disks (Shafee et al.\ 2006) as well as the brightness
profiles of the VLBI images from Sgr~A$^{*}$\ (Broderick \& Loeb 2006;
Noble et al.\ 2007; Dexter et al.\ 2009) will depend on the value of
the quadrupole. Moreover, the location of the ISCO determines the
maximum redshift of relativistically broadened iron
lines. Observations of such lines have been used in the past to infer
the spins of black holes using the Kerr metric (e.g., Brenneman \&
Reynolds 2006; Miller 2007), but are also very sensitive to the
quadrupole moment of the spacetime.

\noindent {\em (ii) The radius of the photon orbit\/}. The radius of
the photon orbit is affected at the same level as the radius of the
ISCO and determines the size of the shadow of the black hole on images
of the accretion flow around it (see Bardeen 1973; Falcke et
al.\ 2000). Interferometric observations of Sgr~A$^*$ in the near
future will allow for a measurement of the detailed structure of the
black-hole shadow in this system (Fish \& Doeleman 2009) and will
place a strong constraint on its quadrupole moment.

\noindent {\em (iii) The velocity of matter at the ISCO\/}. The
accreting material is expected to follow quasi-Keplerian orbits, while
slowly drifting towards the black hole. The high velocity of matter
in these orbits introduces significant Lorentz boosts to the emitted
radiation and causes the approaching region of the flow to appear bluer
and brighter than the receding one. The amount of Lorentz boost
depends on both the dipole and quadrupole moments of the spacetime.
Therefore, measurements of the relative brightness of the blue and red
wings of relativistic broadened lines as well as of the different
regions in the images of accretion flows can be used in constraining
the relative magnitudes of the dipole and quadrupole moments.

\noindent {\em (iv) The detailed trajectories of photons that
  propagate close to the black hole\/}. The amount of gravitational
lensing in the vicinity of the black hole is affected significantly by
the quadrupole moment of the spacetime, as shown in
Figure~\ref{fig:quadrupole}. This leads to non-trivial deformations of
black hole images and of the profiles of relativistically broadened
iron lines, which will be detectable in future high signal-to-noise
observations.

%

One of the main difficulties in constraining the violation of the
no-hair theorem using observations in the electromagnetic spectrum
arises from the existence of degeneracies between changing the dipole
(i.e., the spin of the black hole) and the quadrupole moments of the
spacetime. This would indeed be a problem if we were to use
observables that depend only on the location of the ISCO
(c.f.~Figure~1), such as the maximum temperature of geometrically thin
accretion disks and the maximum redshift of broadened iron
lines. However, the detailed profiles of the continuum and line
spectra as well as of the images from accretion flows depend also very
strongly on the velocity of matter at the ISCO and on the self-lensing
of the radiation emitted near the black hole. As a result, high
signal-to-noise observations encode independent signatures of the
dipole and quadrupole moments of the black-hole metrics that break the
degeneracy between them.

\section{Conclusions}

Astrophysical observations of black holes offer the unique opportunity
of testing the Kerr metric and thus the no-hair theorem, which is one
of the most extreme general relativistic predictions. In this paper,
we presented a framework for such tests using observations of accreting
black holes in the electromagnetic spectrum.

There are at least two types of observations of accreting black holes
that will become possible in the near future and carry the potential
of performing such tests.  First, radio and sub-mm observations of the
black-hole in the center of the Milky Way will be able to produce
snapshots of the innermost accretion flow, resolving the shadow of the
black hole (Doeleman et al.\ 2008; Fish \& Doeleman 2009).  Second,
the broad iron lines and detailed continuum spectra that will be
observed from many accreting black holes with the International X-ray
Observatory will offer an alternative approach to testing the no-hair
theorem.  These observations, in conjunction with the anticipated
detection of gravitational waves with LISA (see Hughes 2006) and the
high-resolution images of the stars in the vicinity of Sgr~A$^*$ (Will
2008), will allow us in the near future to map in detail the
spacetimes of black holes.

\section*{References}

\re
Barausse, E., \& Sotiriou, T.\ 2008, PRL, 101, 9001

\re 
Barcel\'o, C., et al. 2008, PRD, 77, 4032

\re
Bardeen, J.\ 1973, in Black Holes, eds.\ DeWitt \& DeWitt, p.\ 215

\re
Brenneman, L.W., \& Reynolds, C.S., 2006, ApJ, 652, 1028

\re Broderick, A.\,E., Fish, V.\,L., Doeleman, S.\,S., \& Loeb,
A.\ 2009, ApJ, 697, 45

\re
Broderic, A.\,E., \& Loeb, A.\ 2006, ApJ, 636, L109

\re
Collins, N.~A., \& Hughes, S.~A.\ 2004, PRD 69, 124022

\re
Dexter, J., Agol, E., \& Fragile, P.\,C.\ 2009, ApJ, 703, L142

\re
Doeleman, S., et al.\ 2008, Nature, 455, 78

\re
Fabian, A., 2007, in IAU Symposium 238, arXiv:0612435

\re
Falcke, H., Melia, F., \& Agol, E.\ 2000, ApJ, 528, L13

\re
Fish, V.\,L., \& Doeleman, S.\,S.\ 2009, in IAU Symposium 261, in press,
arXiv:0906.4040

\re
Gair, J.\,R., Li, C., \& Mandel, I.\ 2008, PRD, 77, 024035

\re
Glampedakis, K., \& Babak, S.\ 2006, CQG 23, 4167

\re
Geroch, R, 1970, J.\ Math.\ Phys., 11, 2580

\re
Hansen, R.\,O.\ 1974, J.\ Math.\ Phys., 15, 46

\re
Heusler 1996, Black Hole Uniqueness Theorems (Cambridge: University
Press)

\re
Hughes, S.\ 2006, in Sixth International Lisa Symposium, arXiv:gr-qc/0609028

\re
Johannsen, T., \& Psaltis, D., 2009, in preparation

\re
Krolik, J.\,H., \& Hawley, J.\,F.\ 2002, ApJ, 573, 754

\re 
Manko, V.\,S., \& Novikov, I.\,D.\ 1992, CQG 9, 2477

\re
Miller, J.\,M.\ 2007, ARA\&A, 45, 441

\re Nandra, K.\ et al.\ 2006, Astron.\ Nachr., 327, 1039,
arXiv:astro-ph/0610585

\re
Narayan, R., \& McClintock, J.\ 2008, New Astron. Rev.\ 51, 733

\re
Narayan, R., McClintock, J., \& Shafee, R.\ 2007, in Astrophysics of
Compact Objects, arXiv:0710.4073

\re
Noble, S.\,C., et al.\ 2007, CQG 24, S259, arXiv:astro-ph/0701778

\re
Psaltis, D.\ 2004, in X-ray Timing 2003: Rossi and Beyond, arXiv:0402213

\re 
Psaltis, D.\ 2006, in Compact Stellar X-ray sources, eds.\ W.\,H.\,G.\ Lewin
and M.\ van der Klis (Cambridge:University Press)

\re
Psaltis, D.\ 2008, Liv.\ Rev.\ in Relativity, 11, 9, arXiv:0806.1531

\re
Psaltis, D., Perrodin, D., Dienes, K, \& Mocioiu, I.\ 2008, PRL, 100, 1101

\re
Remillard, R.\,A., \& McClintock, J.\,E.\ 2006, ARA\&A, 44, 49

\re
Reynolds, C.\,S., \& Nowak, M.\,A.\ 2003, Phys.\ Rep. 377, 389

\re
Ryan, F.~D.\ 1995, PRD 52, 5707

\re
Scafee, R.\ et al.\ 2006, ApJ, 636, L113

\re
Shapiro, S.\,L., Teukolsky, S.\,A., \& Winicour, J.\ 1995, PRD, 52, 6982

\re
Vigeland, S.\ \& Hughes, S.\ 2009, PRD, submitted
  (arXiv:0911.1756)

\re
Will, C.\,M.\ 2008, ApJ, 674, L25

\re
Yunes, N., \& Pretorius, F.\ 2009, PRD, 79, 4043

\end{document}